\RequirePackage{ifpdf}
\ifpdf % We are running pdfTeX in pdf mode
\documentclass[pdftex]{sigma}
\else
\documentclass{sigma}
\fi

\begin{document}

\numberwithin{equation}{section}

\def\1{\'{\i}}

\def\back{\!\!\!\!\!\!\!\!\!\!}
\def\jp{J_+}
\def\jm{J_-}
\def\jj{J_3}
\def\otra{b}
\def\totra{{\tilde b}}
 \def\kk{K}
\def\kkk{\tilde{K}}
\def\la{\lambda}
 \def\te{\theta}
 \def\d{{\rm d}}
\def\rr{\rho}
\def\tes{\phi}

\def\pois#1#2{\left\{ {#1},{#2} \right\}}
\def\ponm#1#2{\left [ {#1},{#2} \right\}}
\def\co{\Delta}

\def\SW{\rm I}
 \def\Stc{\rm MS}
 \def\cteb{\gamma}
 \def\ctec{\gamma}

\def\pot{{\cal U}}
\def\kc{\rm IKC}

\def\k{\kappa}
\def\>#1{{\mathbf{#1}}}

\allowdisplaybreaks

\renewcommand{\PaperNumber}{026}

\FirstPageHeading

\renewcommand{\thefootnote}{$\star$}

\ShortArticleName{Quantum Deformations and Variable Curvature}

\ArticleName{Quantum Deformations and Superintegrable Motions\\ on
Spaces with Variable Curvature\footnote{This paper is a
contribution to the Proceedings of the O'Raifeartaigh Symposium on
Non-Perturbative and Symmetry Methods in Field Theory (June
22--24, 2006, Budapest, Hungary). The full collection is available
at
\href{http://www.emis.de/journals/SIGMA/LOR2006.html}{http://www.emis.de/journals/SIGMA/LOR2006.html}}}

\Author{Orlando RAGNISCO~$^\dag$,  \'Angel BALLESTEROS~$^\ddag$,
Francisco J.~HERRANZ~$^\ddag$  and Fabio MUSSO~$^\dag\!$}

\AuthorNameForHeading{O. Ragnisco, \'A. Ballesteros, F.J. Herranz
and F. Musso}

\Address{$^\dag$~Dipartimento di Fisica,   Universit\`a di Roma
Tre and Instituto
Nazionale di Fisica Nucleare\\
$\phantom{^\dag}{}$~sezione di Roma Tre,  Via Vasca Navale 84,
I-00146 Roma, Italy}
\EmailD{\href{mailto:ragnisco@fis.uniroma3.it}{ragnisco@fis.uniroma3.it},
\href{mailto:musso@fis.uniroma3.it}{musso@fis.uniroma3.it}}

\Address{$^\ddag$~Departamento de F\1sica, Universidad de Burgos,
E-09001 Burgos, Spain}
\EmailD{\href{mailto:angelb@ubu.es}{angelb@ubu.es},
\href{mailto:fjherranz@ubu.es}{fjherranz@ubu.es}}

\ArticleDates{Received November 12, 2006, in f\/inal form January
22, 2007; Published online February 14, 2007}

\Abstract{An inf\/inite family of quasi-maximally superintegrable
Hamiltonians with a common set of $(2N-3)$ integrals of the motion
is introduced. The integrability properties of all these
Hamiltonians are shown to be a consequence of a hidden
non-standard quantum $sl(2,\mathbb R)$ Poisson coalgebra symmetry.
As a concrete application, one of this Hamiltonians is shown to
 generate the geodesic
 motion on certain manifolds with
a non-constant curvature that turns out to be a function of the
 deformation parameter $z$. Moreover, another Hamiltonian in this family
is shown to generate geodesic motions on Riemannian   and
relativistic spaces all of  whose  sectional curvatures are
constant and equal to the deformation parameter $z$. This approach
can be generalized to arbitrary dimension by making use of
coalgebra symmetry.}

\Keywords{integrable systems; quantum groups; curvature;
contraction; harmonic oscillator; Kepler--Coulomb;  hyperbolic; de
Sitter}

\Classification{37J35; 17B37} % e.g. 35A30; 81Q05
% For 2000 Mathematics Subject Classification see http://www.ams.org/msc/

%%%%%%%%%%%%%%%%%%%%%%%%%%%%%%%%%%%%%%%%%%%%%%%%%%%%%%%%%%%%%%%%%%%%%%%%%%%%%%%%%

\section{Introduction}

The set of known maximally superintegrable systems on the
$N$-dimensional ($N$D) Euclidean space is very limited: it
comprises the isotropic harmonic oscillator with $N$ centrifugal
terms (the so-called Smorodinsky--Winternitz (SW) system
\cite{fris,evans2}), the Kepler--Coulomb (KC) problem with $(N-1)$
centrifugal barriers \cite{MAW} (and some symmetry-breaking
generalizations of it \cite{Pogo}), the
Calogero--Moser--Sutherland model \cite{CMS,SW,RM,Goner1} and some
systems with isochronous potentials~\cite{Gonera}. Both the SW and
the KC systems have  integrals quadratic in the momenta, and also
both of them have been generalized to spaces with non-zero
constant curvature
(see~\cite{VulpiLett,CRMVulpi,RS,PogosClass1,PogosClass2,KalninsH2,Higgs,Leemon,
Schrodingerdual,Schrodingerdualb,kiev}). In order to complete this
brief $N$D summary, Benenti systems on constant curvature spaces
have also to be considered \cite{Benenti}, as well as a maximally
superintegrable deformation of the SW system that was introduced
in \cite{Deform} by making use of quantum algebras.

More recently, the study of 2D and 3D superintegrable systems on
spaces with variable curvature has been
addressed~\cite{darboux1,darboux2,darboux3,darboux4,darboux5,darboux6,darboux7}.
The aim of this paper is to give a~general setting, based on
quantum deformations, for the explicit construction of certain
classes of superintegrable systems on $N$D spaces with
 variable curvature.

In order to f\/ix language conventions, we recall that an
 $N$D completely integrable Hamiltonian~$H^{(N)}$ is called {\it maximally superintegrable} (MS) if there exists
a set of $(2N-2)$ globally def\/ined  functionally independent
constants of the motion that Poisson-commute with~$H^{(N)}$. Among
them, at least two dif\/ferent subsets of $(N-1)$ constants in
involution can be found. In the same way, a system will be called
{\it quasi-maximally superintegrable} (QMS) if there are $(2N-3)$
integrals with the abovementioned properties. All MS systems are
QMS ones, and the latter have only one less integral than the
maximum possible number of functionally independent ones.

In this paper we present the construction of QMS systems on
variable curvature spaces which is just the quantum algebra
generalization of a recent approach to $N$D QMS systems on
constant curvature spaces that include the SW and KC as particular
cases \cite{BHletter}. Some of these variable curvature systems in
2D and 3D have been already studied (see \cite{plb,jpa2D,Checz}),
and we present here the most signif\/icant elements for their $N$D
generalizations. We will show that this scheme is quite
ef\/f\/icient in order to get explicitly a large family of QMS
systems. Among them, some specif\/ic choices for the Hamiltonian
can lead to a MS system, for which only the remaining integral has
to be explicitly found.

In the the next Section we will brief\/ly summarize the $N$D
constant curvature construction given in \cite{BHletter}, that
makes use of an $sl(2,\mathbb R)$ Poisson coalgebra symmetry. The
generic variable curvature approach will be obtained in Section 3
through a non-standard quantum deformation of an $sl(2,\mathbb R)$
Poisson coalgebra. Some explicit 2D and 3D spaces def\/ined
through free motion Hamiltonians will be given in Section 4, and
the $N$D generalization of them will be sketched in Section 5.
Section 6 is devoted to the introduction of some potentials that
generalize the KC and SW ones. A f\/inal Section including some
comments and open questions closes the paper.

%%%%%%%%%%%%%%%%%%%%%%%%%%%%%%%%%%%%%%%%%%%%%%%%%%%%%%%%%%%%%%%%%%%%%%%%%%%%%%%%%

\section[QMS Hamiltonians with  $sl(2,\mathbb R)$ coalgebra symmetry]{QMS Hamiltonians with  $\boldsymbol{sl(2,\mathbb R)}$ coalgebra symmetry}

Let us brief\/ly recall the main result of \cite{BHletter} that
provides an inf\/inite family of QMS Hamiltonians. We stress that,
although some of these Hamiltonians can be interpreted as motions
on spaces with constant curvature, this approach to QMS systems is
quite general, and also non-natural Hamiltonian systems (for
instance, those describing static electromagnetic f\/ields) can be
obtained.

\begin{theorem}[\cite{BHletter}]
Let $\{\>q,\>p \}=\{(q_1,\dots,q_N),(p_1,\dots,p_N)\}$ be $N$
pairs of canonical variables. The ND Hamiltonian
\begin{equation}
H^{(N)}={\cal H}\big(\>q^2,\tilde{\>p}^2,\>q\cdot\>p\big),
\label{hgen}
\end{equation}
with ${\cal H}$ any smooth function and
\begin{equation*}
 \>q^2=\sum_{i=1}^N q_i^2 , \qquad\    \tilde{\>p}^2=
    \sum_{i=1}^N \left(  p_i^2+\frac{\otra_i}{ q_i^2} \right)
\equiv \>p^2 +  \sum_{i=1}^N  \frac{\otra_i}{ q_i^2} ,  \qquad\
\>q\cdot\>p =
  \sum_{i=1}^N  q_i\, p_i ,
%\label{qp}
\end{equation*}
where $b_i$ are arbitrary real parameters, is QMS. The   $(2N-3)$
functionally independent and ``universal" integrals of  motion are
explicitly given by
\begin{gather}
 C^{(m)}= \sum_{1\leq i<j}^m \left\{ ({q_i}{p_j} -
{q_j}{p_i})^2 + \left(
\otra_i\frac{q_j^2}{q_i^2}+\otra_j\frac{q_i^2}{q_j^2}\right)\right\}
+\sum_{i=1}^m \otra_i , \nonumber\\
C_{(m)}= \sum_{N-m+1\leq i<j}^N \left\{ ({q_i}{p_j} -
{q_j}{p_i})^2 + \left(
\otra_i\frac{q_j^2}{q_i^2}+\otra_j\frac{q_i^2}{q_j^2}\right)\right\}
+\sum_{i=N-m+1}^N \otra_i  , \label{cinfm}
\end{gather}
where $m=2,\dots, N$ and $C^{(N)}=C_{(N)}$.  Moreover, the sets of
$N$ functions $\{H^{(N)},C^{(m)}\}$ and $\{H^{(N)},C_{(m)}\}$
$(m=2,\dots, N)$ are in involution.
\end{theorem}

The proof of this general result is based on the observation that,
for any choice of the function ${\cal H}$, the Hamiltonian
$H^{(N)}$ has an $sl(2,\mathbb R)$ Poisson coalgebra
symmetry~\cite{BR} generated by the following Lie--Poisson
brackets and comultiplication map:
\begin{gather}
 \{J_3,J_+\}=2 J_+     ,\qquad
\{J_3,J_-\}=-2 J_- ,\qquad \{J_-,J_+\}=4 J_3  , \label{ba}
\\
\Delta(J_l)=  J_l \otimes 1+ 1\otimes J_l ,\qquad l=+,-,3.
\label{bb}
\end{gather}
The Casimir function for $sl(2,\mathbb R)$ reads
\begin{equation}
{\cal C}=  J_- J_+ -J_3^2  . \label{bc}
\end{equation}

 In fact, the coalgebra approach~\cite{BR} provides an
$N$-particle symplectic realization of $sl(2,\mathbb R)$  through
the $N$-sites coproduct of (\ref{bb}) living on $sl(2,\mathbb
R)\otimes  \cdots^{N)}\otimes sl(2,\mathbb R)$~\cite{Deform}:
\begin{equation}
 J_-=\sum_{i=1}^N q_i^2\equiv \>q^2 ,\qquad    J_+=
    \sum_{i=1}^N \left(  p_i^2+\frac{\otra_i}{ q_i^2} \right)
\equiv \>p^2 +  \sum_{i=1}^N  \frac{\otra_i}{ q_i^2}  ,\qquad J_3=
  \sum_{i=1}^N  q_i p_i\equiv \>q\cdot\>p  ,
\label{be}
\end{equation}
where $\otra_i$ are $N$ arbitrary real parameters. This means that
the $N$-particle generators (\ref{be}) fulf\/il the  commutation
rules (\ref{ba}) with respect to the canonical Poisson bracket. As
a consequence of the coalgebra approach, these generators Poisson
commute with the $(2N-3)$ functions (\ref{cinfm}) given by the
sets $C^{(m)}$ and $C_{(m)}$, which are obtained, in this order,
from the ``left'' and ``right'' $m$-th coproducts of the Casimir
(\ref{bc}) with $m=2,3,\dots,N$ (see \cite{CRMAngel} for details).
Therefore, any   arbitrary function ${\cal H}$ def\/ined on the
$N$-particle symplectic realization of $sl(2,\mathbb R)$
(\ref{be}) is of the form (\ref{hgen}), that is,
\[
{H}^{(N)}= {\cal H}\left(J_-,J_+,J_3\right) ={\cal
H}\left(\>q^2,\>p^2+\sum_{i=1}^N\frac{\otra_i}{q_i^2},\>q\cdot\>p
\right),
\]
 and def\/ines a QMS
Hamiltonian system that Poisson-commutes with all the ``universal
integ\-rals''~$C^{(m)}$ and $C_{(m)}$.

Notice that for arbitrary $N$ there is a single constant of the
motion left  to assure   maximal superintegrability. In this
respect, we   stress that some specif\/ic choices of ${\cal H}$
comprise maximally superintegrable systems as well, but the
remaining integral does not come from the coalgebra symmetry and
has to be deduced by making use of alternative procedures.

Let us now give some explicit examples of this construction.

%%%%%%%%%%%%%%%%%%%%%%%%%%%%%%%%%%%%%%%%%%%%%%%%%%%%%%%%%%%%%%%%%%%
\subsection{Free motion on Riemannian spaces of constant curvature}

It is immediate to realize that the  kinetic energy  ${\cal T}$ of
a particle on the $N$D  Euclidean space ${{\bf E}^N}$ directly
arises through the generator $J_+$ in the symplectic realization
(\ref{be})  with all $\otra_i=0$:
\begin{equation*}
 {\cal H} = {\cal T}=\frac 12 \jp =\frac 12\, {\>p}^2 .
 %\label{bbff}
\end{equation*}

Now the interesting point is that the kinetic energy
  on $N$D Riemannian spaces with constant curvature $\kappa$ can be expressed
in Hamiltonian form as a function of the $N$D symplectic
realization of the $sl(2,\mathbb R)$ generators  (\ref{be}).  In
fact, this can be done in two dif\/ferent ways \cite{BHletter}:
\begin{gather}
\displaystyle{ {\cal H}^{\rm P}={\cal T}^{\rm P}=\frac{1}{2}\left(
1+\k J_-\right)^2 J_+=
\frac{1}{2}\left( 1+\k \>q^2\right)^2 \>p^2} ,\nonumber\\
\displaystyle{ {\cal H}^{\rm B}={\cal T}^{\rm B}=\frac{1}{2}\left(
1+\k J_-\right)\left(  J_+ +\k J_3^2\right)= \frac{1}{2}(1+\k
\>q^2)\left( \>p^2+\k (\>q\cdot \>p)^2 \right) }. \label{dd}
\end{gather}

The function ${\cal H}^{\rm P}$ is just the kinetic energy for a
free particle  on  the spherical ${\bf S}^N$ ($\k>0$) and
hyperbolic ${\bf H}^N$    ($\k<0$) spaces  when this is expressed
in terms of Poincar\'e coordinates $\>q$ and canonical momenta
$\>p$ (coming from a stereographic projection in ${\mathbb
R^{N+1}}$);  on the other hand ${\cal H}^{\rm B}$ corresponds to
Beltrami coordinates and momenta (central projection). By
construction, both Hamiltonians are QMS ones since they
Poisson-commute with the integrals (\ref{cinfm}).

%%%%%%%%%%%%%%%%%%%%%%%%%%%%%%%%%%%%%%%%%%%%%%%%%%%%%%%%%%%%%%%%%%%%%%%%%%%%%%%%%

\subsection{Superintegrable potentials on Riemannian spaces of constant curvature}

QMS potentials ${\cal V}$ on constant curvature spaces can now be
constructed by adding some suitable functions depending on $J_-$
to (\ref{dd}) and by considering arbitrary centrifugal  terms that
come from symplectic realizations of the $J_+$ generator with
generic $\otra_i$'s:
\begin{equation*}
{\cal H}={\cal T}(J_+,J_-,J_3) + {\cal V}(J_-).
\end{equation*}
The Hamiltonians that we will obtain in this way are the curved
counterpart of the Euclidean systems, and through dif\/ferent
values of the curvature $\k$ we will simultaneously cover the
cases~${\bf S}^N$ ($\k>0$), ${\bf H}^N$    ($\k<0$), and ${\bf
E}^N$
 ($\k=0$).

In order to motivate the choice of the potential functions ${\cal
V}(J_-)$, it is important to recall that in the constant curvature
analogues of the oscillator and KC problems the Euclidean radial
distance $r$ is just replaced by the function $
\frac{1}{\sqrt{\k}}\tan(\sqrt{\k}\,r) $ (see \cite{BHletter} for
the expression of this quantity in terms of Poincar\'e and
Beltrami coordinates). Also, for the sake of simplicity, the
centrifugal terms coming from the symplectic realization with
arbitrary $\otra_i$ will be expressed in ambient coordinates
$x_i$~\cite{BHletter}:
\[
{\mbox{Poincar\'e:}}\quad  x_i=  \frac{2 q_i}{1+ \k\>q^2};\qquad
 {\mbox{Beltrami:}} \quad
x_i=\frac{q_i}{\sqrt{1+ \k \>q^2}}.
\]

Special  choices for ${\cal V}(J_-)$ lead to the following
systems, that are always expressed in both Poincar\'e and Beltrami
phase spaces:
\medskip

\noindent $\bullet$ {\em A curved Evans system}. The constant
curvature generalization of a 3D Euclidean system with radial
symmetry~\cite{Evansa} would be given by
\begin{gather}
\displaystyle{ {\cal H}^{\rm P}={\cal T}^{\rm P}+{\cal V} \left(
\frac{4 J_-}{ (1-\k J_-)^2 } \right) = \frac12 {\left( 1+\k
\>q^2\right)^2 \>p^2}}+{\cal V}\left( \frac{4\>q^2}{(1-\k\>q^2)^2}
\right) + \sum_{i=1}^N\frac{2\otra_i}{x_i^2},
\nonumber\\
\displaystyle{ {\cal H}^{\rm B}={\cal T}^{\rm B}+ {\cal
V}\left(J_-\right)= \frac{1}{2}(1+\k \>q^2)\left( \>p^2+\k
(\>q\cdot \>p)^2 \right) +{\cal V}\left(
\>q^2\right)+\sum_{i=1}^N\frac{\otra_i}{2 x_i^2}}, \label{eb}
\end{gather}
where ${\cal V}$ is an arbitrary smooth function that determines
the  central potential;   the specif\/ic  dependence  on $J_-$ of
${\cal V}$ corresponds to the square of the radial distance in
each coordinate system.

\medskip
\noindent $\bullet$ {\em The curved Smorodinsky--Winternitz
 system}~\cite{VulpiLett,CRMVulpi,RS,PogosClass1,PogosClass2,KalninsH2}.
Such a system is just the Higgs oscillator~\cite{Higgs,Leemon}
with angular frequency $\omega$ (that arises as the argument of
${\cal V}$ in (\ref{eb})) plus the corresponding centrifugal
terms:
\begin{gather*}
 {\cal H}^{\rm P}={\cal T}^{\rm P}+   \frac{4 \omega^2
J_-}{ (1-\k J_-)^2 } = \frac12 {\left( 1+\k \>q^2\right)^2 \>p^2}+
\frac{4\omega^2 \>q^2}{(1-\k\>q^2)^2}   +
\sum_{i=1}^N\frac{2\otra_i}{x_i^2} ,
\nonumber\\
{\cal H}^{\rm B}={\cal T}^{\rm B}+\omega^2 J_- = \frac{1}{2}(1+\k
\>q^2)\left( \>p^2+\k (\>q\cdot \>p)^2 \right) + \omega^2
\>q^2+\sum_{i=1}^N\frac{\otra_i}{2 x_i^2}.
%\label{ec}
\end{gather*}
This is a     MS Hamiltonian and the remaining constant of the
motion can be chosen from any of  the following $N$ functions:
\begin{gather*}
{\cal I}_i^{\rm P}=\left( p_i(1-\k \>q^2) + 2\k (\>q\cdot \>p) q_i
\right)^2+ \frac{8  \omega^2 q_i^2}{(1-\k\>q^2)^2}   +
{\otra_i}\,
\frac{(1-\k\>q^2)^2} { q_i^2}  ,\nonumber\\
 {\cal I}_i^{\rm B}=\left( p_i+\k (\>q\cdot \>p) q_i \right)^2+2
\omega^2  q_i^2+  {\otra_i}/{q_i^2},\qquad i=1,\dots,N.
%\label{cccx}
\end{gather*}

\noindent $\bullet$ {\em A  curved generalized Kepler--Coulomb
system}~\cite{RS, PogosClass1,PogosClass2,Schrodingerdual,
Schrodingerdualb,kiev}. The     curved KC potential with real
constant $k$ together with $N$ centrifugal terms would be given by
\begin{gather*}  {\cal H}^{\rm P}={\cal T}^{\rm P}-k \left(\frac{4
J_-}{ (1-\k J_-)^2 } \right)^{-1/2}= \frac12 {\left( 1+\k
\>q^2\right)^2 \>p^2}- k \, \frac{(1-\k\>q^2)}   {2
\sqrt{\>q^2}}+ \sum_{i=1}^N\frac{2\otra_i}{x_i^2},
\nonumber\\
{\cal H}^{\rm B}={\cal T}^{\rm B} -k  J_-^{-1/2}= \frac{1}{2}(1+\k
\>q^2)\left( \>p^2+\k (\>q\cdot \>p)^2 \right) -\frac{k}{
\sqrt{\>q^2 }}+\sum_{i=1}^N\frac{\otra_i}{2 x_i^2} .
%\label{ee}
\end{gather*}
This is again  a MS system provided that, at least, one
$\otra_i=0$. In this case the remaining   constant of the motion
turns out to be
\begin{gather}
 {\cal L}_i^{\rm P}=\sum_{l=1 }^N \left(
p_l(1-\k \>q^2) + 2\k (\>q\cdot \>p) q_l \right) (q_l p_i-q_i p_l)
+ \frac{k  q_i}{2\sqrt{\>q^2}}-
\sum_{l=1;l\ne i}^N \otra_l\,\frac{ q_i(1-\k \>q^2)}{ q_l^2}   ,\nonumber\\
 {\cal L}_i^{\rm B}=\sum_{l=1 }^N  \left(
p_l+\k (\>q\cdot \>p) q_l \right)   (q_l p_i-q_i p_l)
 +\frac{ k q_i}{\sqrt{\>q^2}}-
\sum_{l=1;l\ne i}^N \otra_l\,\frac{ q_i}{ q_l^2}  . \label{ceex}
\end{gather}
If another $\otra_j=0$, then ${\cal L}_j^{\rm P,B}$ is also a new
constant of the motion. In this way the proper curved KC
system~\cite{Schrodinger} (with   all the $\otra_i$'s equal to
zero) is obtained, and in that case (\ref{ceex}) are just
the~$N$~components of the  Laplace--Runge--Lenz vector on ${\bf
S}^N$ $(\k>0)$ and ${\bf H}^N$ $(\k<0)$.

We also stress that all these examples share the {\em same} set of
constants of the motion  (\ref{cinfm}), although the geometric
meaning of the canonical coordinates and momenta can be
dif\/ferent.

%%%%%%%%%%%%%%%%%%%%%%%%%%%%%%%%%%%%%%%%%%%%%%%%%%%%%%%%%%%%%%%%%%%
\section[QMS Hamiltonians with quantum deformed $sl(2,\mathbb R)$ coalgebra symmetry]{QMS Hamiltonians with quantum deformed $\boldsymbol{sl(2,\mathbb R)}$ coalgebra\\ symmetry}

Here  we will show   that a generalization of the construction
presented in the previous Section  can be obtained through a
quantum deformation of   $sl(2,\mathbb R)$,  yielding  QMS systems
for
 certain spaces with variable curvature. Let us now state the general
statement that provides a superintegrable deformation of Theorem
1.

\begin{theorem}
Let $\{\>q,\>p \}=\{(q_1,\dots,q_N),(p_1,\dots,p_N)\}$ be $N$
pairs of canonical variables. The N{\rm{D}} Hamiltonian
\begin{equation}
H^{(N)}_z={\cal
H}_z\left(\>q^2,\tilde{\>p}_z^2,(\>q\cdot\>p)_z\right),
\label{hgenb}
\end{equation}
where ${\cal H}_z$ is any smooth function and
\begin{gather*}
\>q^2= \sum_{i=1}^N q_i^2,\qquad
 \tilde{\>p}_z^2=\sum_{i=1}^N
\left( \frac {\sinh z q_i^2}{z q_i^2} \, p_i^2  +\frac{z
\otra_i}{\sinh z
q_i^2} \right) {\rm e}^{z \kk_i^{(N)}(q^2) },\\%\label{sympz}\\
(\>q\cdot\>p)_z=\sum_{i=1}^N \frac {\sinh z q_i^2}{z q_i^2} \,
q_ip_i \, {\rm e}^{z \kk_i^{(N)}(q^2) }  , \nonumber
\end{gather*}
with
\begin{equation}
  \kk_i^{(h)}(q^2)=  - \sum_{k=1}^{i-1}  q^2_k+
\sum_{l=i+1}^h   q^2_l , \label{kki}
\end{equation}
is QMS for any choice of the function ${\cal H}$ and for arbitrary
real parameters $b_i$. The $(2N-3)$ functionally independent and
``universal" integrals of the motion
   are   given by
\begin{gather}
C_z^{(m)}= \sum_{1\leq i<j}^m{Q_{ij}^z}\,{\rm e}^{ z
\kk_{ij}^{(m)}(q^2)} + \sum_{i=1}^m \otra_i \, {\rm e}^{2 z
\kk_{i}^{(m)}(q^2)},\nonumber\\
C_{z,(m)}= \sum_{N-m+1\leq i<j}^N{Q_{ij}^z}\,{\rm e}^{ z
\kkk_{ij}^{(N-m+1)}(q^2)} + \sum_{i=N-m+1}^N \otra_i\,  {\rm e}^{2
z \kkk_{i}^{(N-m+1)}(q^2)}, \label{fk}
\end{gather}
where  $m=2,\dots, N$,  $C_z^{(N)}=C_{z,(N)}$, and
\begin{gather*}
 \kk_{ij}^{(h)}(q^2)  = \kk_i^{(h)}(q^2)  +  \kk_j^{(h)}(q^2)
  =    - 2\sum_{k=1}^{i-1}   q^2_k  -    q^2_i  +    q^2_j  +
2\sum_{l=j+1}^h   q^2_l  ,\nonumber\\
\kkk_i^{(h)}(q^2)=  - \sum_{k=h}^{i-1}  q^2_k+
\sum_{l=i+1}^N   q^2_l,\nonumber\\
 \kkk_{ij}^{(h)}(q^2)  = \kkk_i^{(h)}(q^2)  +  \kkk_j^{(h)}(q^2)
  =    - 2\sum_{k=h}^{i-1}   q^2_k  -    q^2_i  +    q^2_j  +
2\sum_{l=j+1}^N   q^2_l, \nonumber\\
Q_{ij}^z=\left\{\frac {\sinh z q_i^2}{z  q_i^2}\, \frac {\sinh z
q_j^2}{z q_j^2} \left({q_i}{p_j} - {q_j}{p_i}\right)^2   +\left(
\otra_i\, \frac {\sinh z q_j^2}{\sinh z q_i^2} + \otra_j\, \frac
{\sinh z q_i^2}{\sinh z q_j^2} \right)\right\},
\end{gather*}
with $i<j$.  Moreover, the sets of $N$ functions
$\{H^{(N)}_z,C^{(m)}_z\}$ and $\{H^{(N)}_z,C_{z,(m)}\}$
$(m=2,\dots, N)$ are in involution.
\end{theorem}

%%%%%%%%%%%%%%%%%%%%%%%%%%%%%%%%%%%%%%%%%%%%%%%%%%%%%%%%%%%%%%%%%%%%%%%%%%%%%%%%%

\subsection{The proof}

The proof  is based on the fact that, for any choice of the
function ${\cal H}$, the Hamiltonian $H^{(N)}_z$ has a~deformed
Poisson coalgebra symmetry, $sl_z(2,\mathbb R)$, coming (under a
certain   symplectic realization)  from the  non-standard quantum
deformation of    $sl(2,\mathbb R)$~\cite{Ohn,CP} where $z$ is the
deformation parameter $(q={\rm e}^z)$. If we perform  the   limit
$z\to 0$ in all the results given in Theorem 2, we shall exactly
recover Theorem 1.
  Here we
sketch the main steps of this construction, referring   to
\cite{Deform,CRMAngel} for further details.

We recall that the non-standard $sl_z(2,\mathbb R)$ Poisson
coalgebra is given by the following deformed Poisson brackets and
coproduct~\cite{Deform}:
\begin{gather}
\{\jj,\jp\}=2 \jp \cosh z\jm  ,\qquad
 \{\jj,\jm\}=-2\,\frac {\sinh z\jm}{z} ,\qquad
 \{\jm,\jp\}=4 \jj ,
\label{baa}
\\
  \Delta_z(\jm)=  \jm \otimes 1+
1 ,\qquad  \Delta_z(J_l)=J_l \otimes {\rm e}^{z \jm} + {\rm e}^{-z
\jm} \otimes J_l ,\qquad l=+,3. \label{ggcc}
\end{gather}
The Casimir function for $sl_z(2,\mathbb R)$ reads
\begin{equation}
{\cal C}_z=  \frac {\sinh z\jm}{z}\, \jp -\jj^2 . \label{gc}
\end{equation}

A  one-particle symplectic realization of (\ref{baa}) is given by
\begin{equation*}
\jm^{(1)}=q_1^2  ,\qquad
 \jp^{(1)}=\frac {\sinh z q_1^2}{z q_1^2}\, p_1^2 +
\frac{z \otra_1}{\sinh z q_1^2},\qquad \jj^{(1)}=\frac {\sinh z
q_1^2}{z q_1^2}\, q_1 p_1 ,
\end{equation*}
where $\otra_1$ is a real parameter that labels the representation
through ${\cal C}_z=\otra_1$.

Now the essential point is the fact that the coalgebra
approach~\cite{BR} provides the corresponding $N$-particle
symplectic realization of $sl_z(2,\mathbb R)$  through the
$N$-sites coproduct of (\ref{ggcc}) living on $sl_z(2,\mathbb
R)\otimes  \cdots^{N)}\otimes sl_z(2,\mathbb R)$~\cite{Deform}:
\begin{gather}
\jm^{(N)}= \sum_{i=1}^N q_i^2 \equiv \>q^2 ,\qquad
\jj^{(N)}=\sum_{i=1}^N \frac {\sinh z q_i^2}{z q_i^2} \, q_ip_i \,
{\rm e}^{z \kk_i^{(N)}(q^2) }
\equiv (\>q\cdot\>p)_z ,\nonumber \\
 \jp^{(N)}=\sum_{i=1}^N
\left( \frac {\sinh z q_i^2}{z q_i^2} \, p_i^2  +\frac{z
\otra_i}{\sinh z q_i^2} \right)  {\rm e}^{z \kk_i^{(N)}(q^2)
}\equiv \tilde{\>p}_z^2 , \label{zsymp}
\end{gather}
where $\kk_i^{(N)}(q^2)$ is def\/ined in (\ref{kki}) and $\otra_i$
are $N$ arbitrary real parameters that label the representation on
each ``lattice'' site. This means that the $N$-particle generators
(\ref{zsymp}) fulf\/il the  commutation rules (\ref{baa}) with
respect to the   canonical Poisson bracket
\[
\{f,g\}=\sum_{i=1}^N\left(\frac{\partial f}{\partial q_i}
\frac{\partial g}{\partial p_i} -\frac{\partial g}{\partial q_i}
\frac{\partial f}{\partial p_i}\right).
%\label{bbgg}
\]
Therefore the Hamiltonian  (\ref{hgenb}) is obtained through an
arbitrary smooth function
  ${\cal H}_z$ def\/ined on the
$N$-particle symplectic realization of the generators of
$sl_z(2,\mathbb R)$:
\begin{equation}
{H}^{(N)}_z= {\cal H}_z\big(\jm^{(N)},\jp^{(N)},\jj^{(N)}\big)
={\cal H}_z\left(\>q^2,\tilde{\>p}_z^2,(\>q\cdot\>p)_z \right).
\label{hamham}
\end{equation}

By construction~\cite{BR}, the functions (\ref{zsymp}) Poisson
commute with the $(2N-3)$ functions~(\ref{fk}) given by the sets
$C^{(m)}_z$ and $C_{z,(m)}$, which are obtained from the ``left"
and ``right" $m$-th copro\-ducts of the Casimir (\ref{gc}) with
$m=2,3,\dots,N$~\cite{CRMAngel}. For instance, the $C_z^{(m)}$
integrals are nothing but
\begin{equation*}
C_z^{(m)}= \frac {\sinh z\jm^{(m)}}{z}\, \jp^{(m)} -
\big(\jj^{(m)}\big)^2  ,
\end{equation*}
and the right ones $C_{z,(m)}$ can be obtained through an
appropriate permutation of the labelling of the lattice sites
(note that these integrals depend on the canonical coordinates
running from $(N-m+1)$ up to $N$). Thus ${H}^{(N)}_z$  Poisson
commutes with the $(2N-3)$ integrals and, furthermore, the
coalgebra symmetry also ensures that each of the subsets
$\{C_z^{(2)}, \dots ,C_z^{(N)},{H}^{(N)}_z \}$ and
$\{C_{z,(2)},\dots ,C_{z,(N)},{H}^{(N)}_z \}$ consists of  $N$
functions in involution.

In order to prove the functional independence of the   $2N-2$
functions $\{ C_z^{(2)},C_z^{(3)},\dots ,C_z^{(N)}\equiv
C_{z,(N)}, C_{z,(N-1)},\dots ,C_{z,(2)},{H}^{(N)}_z \}$ it
suf\/f\/ices to realize that such functions are just deformations
in the deformation parameter $z$ of the $sl(2,\mathbb R)$
integrals given by (\ref{cinfm}), and the latter (which are
recovered when $z\to 0$) are indeed functionally independent.

Thus, we  conclude that any   arbitrary function ${\cal H}_z$
(\ref{hamham}) def\/ines a QMS Hamiltonian system.

%%%%%%%%%%%%%%%%%%%%%%%%%%%%%%%%%%%%%%%%%%%%%%%%%%%%%%%%%%%%%%%%%%%%%%%%%%%%%%%%%

\subsection[The $N=2$ case]{The $\boldsymbol{N=2}$ case}

In order to illustrate the previous construction, let us
explicitly write the $2$-particle symplectic realization of
$sl_z(2,\mathbb R)$ (\ref{zsymp}):
\begin{gather*}
 \jm^{(2)}=q_1^2+q_2^2  , \qquad \jj^{(2)}=
 \frac {\sinh z q_1^2}{z q_1^2 } \, {\rm e}^{z q_2^2}   q_1 p_1  +
 \frac {\sinh z q_2^2}{z q_2^2 }\, {\rm e}^{-z q_1^2}    q_2 p_2  ,\nonumber\\
 \jp^{(2)}=
  \frac {\sinh z q_1^2}{z q_1^2}\,  {\rm e}^{z q_2^2}   p_1^2   +
 \frac {\sinh z q_2^2}{z q_2^2} \,  {\rm e}^{-z q_1^2}   p_2^2 +\frac{z
\otra_1}{\sinh z q_1^2}\,{\rm e}^{z q_2^2} + \frac{z
\otra_2}{\sinh z q_2^2} \,  {\rm e}^{-z q_1^2} .
%\label{sympz2}
\end{gather*}
In this case there is a single (left and right) constant of the
motion:
\begin{equation*}
C_z^{(2)}= \frac {\sinh z\jm^{(2)}}{z}\, \jp^{(2)} -
\big(\jj^{(2)}\big)^2  .
\end{equation*}
After some straightforward computations this integral can be
expressed as
\begin{gather}
   C^{(2)}_z=
\frac {\sinh z q_1^2}{z  q_1^2}\, \frac {\sinh z q_2^2}{z q_2^2}
\left({q_1}{p_2} - {q_2}{p_1}\right)^2  {\rm e}^{ z (q_2^2 -
q_1^2)} +  \otra_1  {\rm e}^{2 z q_2^2 }+  \otra_2  {\rm e}^{-2 z
q_1^2 }\cr \phantom{C^{(2)}_z=}{}+ \left( \otra_1 \frac {\sinh z
q_2^2}{\sinh z q_1^2} + \otra_2 \frac {\sinh z q_1^2}{\sinh z
q_2^2} \right)  {\rm e}^{ z (q_2^2 - q_1^2)} . \label{casdos}
\end{gather}
By construction, this constant of the motion will Poisson-commute
with all the Hamiltonians
\[
{H}^{(2)}_z= {\cal H}_z\left(\jm^{(2)},\jp^{(2)},\jj^{(2)}\right).
\]

Note that in the $N=2$ case quasi-maximal superintegrability means
only integrability, i.e., the only constant given by   Theorem 2
is just $C_z^{(2)}\equiv C_{z,(2)}$; this fact does not exclude
that there could be some specif\/ic choices for ${\cal H}_z$ for
which an additional integral does exist. When $N\ge 3$, Theorem 2
will always provide QMS Hamiltonians.

%%%%%%%%%%%%%%%%%%%%%%%%%%%%%%%%%%%%%%%%%%%%%%%%%%%%%%%%%%%%%%%%%%%
\section{Free motion on 2D and 3D curved manifolds}

%%%%%%%%%%%%%%%%%%%%%%%%%%%%%%%%%%%%%%%%%%%%%%%%%%%%%%%%%%%%%%%%%%%
\subsection{2D curved manifolds}

Throughout this Section we will consider {only free motion}.
Therefore we shall take the symplectic realization with
$\otra_1=\otra_2=0$ in order to avoid centrifugal potential terms.
In general, we can consider an inf\/inite family of {\em
integrable} (and quadratic
 in the momenta) free $N=2$ motions with $sl_z(2,\mathbb R)$ coalgebra symmetry
through Hamiltonians of the type
\begin{equation}
 {H}^{(2)}_z =\frac 12 \jp^{(2)} f\bigl(z\jm^{(2)}\bigr),
 \label{bbfff}
\end{equation}
where $f$ is an arbitrary smooth function such that
 $\displaystyle{\lim_{z\to 0}f\bigl(z\jm^{(2)}\bigr)=1}$, that is,
$\lim\limits_{z\to 0}{H}^{(2)}_z=\frac 12 (p_1^2+p_2^2)$.  We
shall explore in the sequel some specif\/ic choices for $f$, and
we shall analyse the spaces generated by them.

%%%%%%%%%%%%%%%%%%%%%%%%%%%%%%%%%%%%%%%%%%%%%%%%%%%%%%%%%%%%%%%%%%%%%%%%%%%%%%%%%

\subsubsection{An integrable case}

Of course, the
 simplest choice will be just to set $f\equiv 1$ \cite{plb}:
\begin{equation}
{\cal H}^{\rm \SW}_z=\frac 12
 \jp^{(2)}=\frac12 \left( \frac {\sinh z
 q_1^2}{z q_1^2} \, {\rm e}^{z q_2^2}  p_1^2   +
 \frac {\sinh z q_2^2}{z q_2^2} \, {\rm e}^{-z q_1^2}  p_2^2  \right)   .
 \label{bg}
\end{equation}
Hence the {kinetic energy} ${\cal T}^{\rm \SW}_z(q_i,p_i)$ coming
from
 ${\cal H}^{\rm \SW}_z$  is
\begin{equation}
 {\cal T}^{\rm \SW}_z(q_i,\dot q_i)=\frac 12 \left(\frac
 {z q_1^2}{\sinh z q_1^2} \, {\rm e}^{-z q_2^2} \dot q_1^2   +
 \frac {z q_2^2}{\sinh z q_2^2} \, {\rm e}^{z q_1^2} \dot q_2^2
 \right),
 \label{ca}
\end{equation}
 and def\/ines a geodesic f\/low on a 2D
 Riemannian space with signature
 diag$(+,+)$ and metric given by:
\begin{equation}
 \d s_I^2=\frac {2z q_1^2}{\sinh z
 q_1^2} \, {\rm e}^{-z q_2^2} \,\d q_1^2   +
  \frac {2 z q_2^2}{\sinh z q_2^2} \, {\rm e}^{z q_1^2}\, \d q_2^2  .
 \label{cc}
\end{equation}
 The Gaussian curvature $K$ for this space can be computed through
\begin{equation*}
 K=\frac{-1}{\sqrt{g_{11} g_{22}}}\left\{ \frac{\partial}{\partial
 q_1} \left( \frac{1}{\sqrt{g_{11}}} \frac{\partial
 \sqrt{g_{22}}}{\partial q_1}
 \right)+
 \frac{\partial}{\partial
 q_2} \left( \frac{1}{\sqrt{g_{22}}} \frac{\partial
 \sqrt{g_{11}}}{\partial q_2}
 \right)\right\} ,
\end{equation*}
and turns out to be {non-constant and {\em negative}}:
\begin{equation*}
 K(q_1,q_2;z)=-   z \sinh\left(z(q_1^2+q_2^2) \right).
 %\label{cd}
\end{equation*}
Therefore, the
 underlying 2D space is of {hyperbolic  type} and
 endowed with  a {``radial" symmetry}.

Let us now consider the following change of coordinates that
includes a new parameter $\la_2\ne 0$:
\begin{gather*}
 \cosh(\la_1 \rr)=\exp\left\{z(q_1^2+q_2^2)\right\},\qquad
  \sin^2(\la_2 \te)=\frac{\exp\left\{2z q_1^2
 \right\}-1}{\exp\left\{2z(q_1^2+q_2^2)\right\}-1},
\end{gather*}
 where $z=\lambda_1^2$  and $\la_2$ can take either a real or a
 pure imaginary value.
Note that the new variable $ \cosh (\lambda_1 \rho)$ is a
collective variable, a function of $\Delta(J_-)$; its role will be
specif\/ied later. On the other hand, the zero-deformation limit
$z\to 0$  is in fact the f\/lat limit $K \to 0$, since in this
limit
\begin{gather*}
\rho \to 2(q_1^2+q_2^2), \qquad \sin ^2 (\lambda_2 \theta) \to
\frac{q_1^2}{q_1^2 + q_2^2}.
\end{gather*}
Thus $\rr$ can be interpreted as  a {radial coordinate} and $\te$
 is  {either a circular ($\la_2$ real) or a hyperbolic angle}  ($\la_2$ imaginary). Notice
that in the latter case, say $\la_2={\rm i}$, the coordinate $q_1$
is imaginary and can be written as $q_1={\rm i} \tilde q_1$ where
$\tilde q_1$ is a real coordinate; then $\rho \to 2(q_2^2-\tilde
q_1^2)$ which corresponds to a relativistic radial distance.
Therefore the introduction of the additional parameter $\la_2$
will allow us to obtain Lorentzian metrics.

 In
this new coordinates, the metric (\ref{cc}) reads
\begin{equation*}
 \d s_I^2=\frac {1}{\cosh(\la_1 \rr)}
 \left( \d \rr^2  +\la_2^2\,\frac{\sinh^2(\la_1 \rr)}{\la_1^2} \, \d
 \te^2  \right) =\frac {1}{\cosh(\la_1 \rr)}\,\d s_0^2,
 \label{cg}
\end{equation*}
where $\d s_0^2$ is just the metric of the 2D Cayley--Klein spaces
 in terms of geodesic polar coordinates~\cite{ramon,Conf}
  provided that we identify $z=\la_1^2\equiv -\kappa_1$ and
 $\la_2^2\equiv \kappa_2$; hence $\la_2$  determines
the signature of the metric. The {Gaussian curvature} turns out to
be
\begin{equation*}
 K(\rr)=-\frac 12 \la_1^2 \,\frac{\sinh^2(\la_1 \rr)}{\cosh(\la_1
 \rr)} .
 %\label{cj}
\end{equation*}
In this way we  f\/ind the {following   spaces}, whose main
properties are summarized in Table 1:

\begin{itemize}\itemsep=0pt

 \item[$\bullet$]  When  $\la_2$ is real, we get a  2D {deformed
 sphere} ${\bf S}^2_z$ $(z<0)$,    and  a {deformed
 hyperbolic or  Lobachewski space} ${\bf H}^2_z$ $(z>0)$.

 \item[$\bullet$]  When  $\la_2$ is imaginary, we obtain a {deformation
 of   the (1+1)D   anti-de Sitter spacetime} ${\bf AdS}_z^{1+1}$ $(z<0)$ and
of the   {de Sitter one} ${\bf dS}_z^{1+1}$
 $(z>0)$.

 \item[$\bullet$] In the non-deformed case $z\to 0$,  the Euclidean
 space
 ${\bf E}^2$
 ($\la_2$ real) and  Minkowskian spacetime
 ${\bf M}^{1+1}$ ($\la_2$ imaginary) are recovered.

 \end{itemize}

%%%%%%%%%%%%%%%%%%%% table1 %%%%%%%%%%%%%%%%%%%%%%
 \begin{table}
 {\footnotesize
  \noindent
 \caption{{{Metric and Gaussian curvature of the 2D spaces with $sl_z(2,\mathbb R)$
coalgebra symmetry for
 dif\/ferent values of the deformation parameter $z=\lambda_1^2$ and
 signature parameter $\lambda_2$.}}}
 \label{table1}
 \medskip
 \noindent\hfill
 $$
\back\back  \begin{array}{ll}
 \hline
 \\[-6pt]
 {\mbox {2D deformed Riemannian spaces}}&\quad{\mbox  {$(1+1)$D
 deformed relativistic spacetimes}}\\
 [4pt]
 \hline
 \\
 [-6pt]
 \mbox {$\bullet$ {Deformed sphere} ${\bf S}^2_z$}&\quad\mbox
{$\bullet$
 {Deformed anti-de Sitter spacetime} ${\bf AdS}^{1+1}_z$}\\[4pt] z=-1;\
 (\la_1,\la_2)=({\rm i},1)&\quad z=-1;\ (\la_1,\la_2)=({\rm i},{\rm
 i})\\[4pt]
 \displaystyle{\d s^2 =\frac{1}{\cos \rr}\left( \d \rr^2+\sin^2
 \rr\,\d\te^2
 \right)} &\quad
 \displaystyle{\d s^2 =\frac{1}{\cos \rr}\left( \d \rr^2-\sin^2
 \rr\,\d\te^2
 \right)} \\[8pt]
  \displaystyle{K =-\frac{\sin^2 \rr}{2\cos \rr} } &\quad
  \displaystyle{K =-\frac{\sin^2 \rr}{2\cos \rr} } \\[12pt]
 \mbox {$\bullet$ Euclidean space  ${\bf E}^2$}&\quad\mbox {$\bullet$
 Minkowskian spacetime ${\bf M}^{1+1}$}\\[4pt]
 z=0;\ (\la_1,\la_2)=(0,1)&\quad
 z=0;\ (\la_1,\la_2)=(0,{\rm i})\\[4pt]
  \displaystyle{\d s^2 =  \d \rr^2+ \rr^2\d\te^2
  } &\quad
  \displaystyle{\d s^2 =  \d \rr^2- \rr^2\d\te^2} \\[2pt]
  \displaystyle{K =0 } &\quad
  \displaystyle{K =0} \\[6pt]
 \mbox {$\bullet$ {Deformed hyperbolic space} ${\bf H}_z^2$}&\quad\mbox
 {$\bullet$ {Deformed de Sitter spacetime} ${\bf dS}^{1+1}_z$}\\[4pt]
 z=1;\ (\la_1,\la_2)=(1,1)&\quad
 z=1;\ (\la_1,\la_2)=(1,{\rm i})\\[4pt]
 \displaystyle{\d s^2 =\frac{1}{\cosh \rr}\left( \d \rr^2+\sinh^2
 \rr\,\d\te^2
 \right)} &\quad
 \displaystyle{\d s^2 =\frac{1}{\cosh \rr}\left( \d \rr^2-\sinh^2
 \rr\,\d\te^2
 \right)} \\[8pt]
 \displaystyle{K =-\frac{\sinh^2 \rr}{2\cosh \rr} } &\quad
  \displaystyle{K =-\frac{\sinh^2 \rr}{2\cosh \rr} } \\[8pt]
 \hline
 \end{array}
 $$
 \hfill}
 \end{table}
 %%%%%%%%%%%%%%%%%%%%%%%%%%%%%%%%%%%%%%%%%%%%%%%%%%%%%%%%%%%%%%%%%%%%

\medskip
\noindent Accordingly, the kinetic energy  (\ref{ca}) is
transformed into
\begin{equation*}
 {\cal T}^{\rm \SW}_z(\rr,\te;\dot \rr,\dot \te)=\frac
 {1}{2\cosh(\la_1 \rr)}
 \left(\dot \rr^2  +\la_2^2\,\frac{\sinh^2(\la_1 \rr)}{\la_1^2} \,
 \dot \te^2
 \right),
\end{equation*}
and  the free motion Hamiltonian (\ref{bg}) is written as
\begin{equation*}
 \widetilde{H}^{\SW}_z=\frac 12 \cosh(\la_1
 \rr)\left(p_\rr^2 +\frac{\la_1^2}{\la_2^2\sinh^2(\la_1 \rr)} \,
 p_\te^2\right),
 %\label{dda}
\end{equation*}
where $\widetilde{H}^{\SW}_z =2 {\cal H}^{\SW}_z$. There is a
{unique} constant of the motion ${ C}^{(2)}_z\equiv{ C}_{z,(2)}$
(\ref{casdos}) which in terms of the new phase space is simply
given by
\begin{equation*}
  \widetilde{C}_z=p_\te^2 ,
\end{equation*}
provided that $\widetilde{C}_z= 4\la_2^2  { C}^{(2)}_z$.
 This allows us to apply a   radial-symmetry reduction:
\begin{equation*}
 \widetilde{H}^{\SW}_z =\frac 12 \cosh(\la_1
 \rr)\, p_\rr^2  +\frac{\la_1^2 \cosh(\la_1
 \rr)}{2\la_2^2\sinh^2(\la_1 \rr)} \, \widetilde{C}_z.
 %\label{dede}
\end{equation*}
We remark that the explicit integration of the geodesic motion on
all these spaces can be explicitly performed in terms of elliptic
integrals.

%%%%%%%%%%%%%%%%%%%%%%%%%%%%%%%%%%%%%%%%%%%%%%%%%%%%%%%%%%%%%%%%%%%%%%%%%%%%%%%%%

\subsubsection{The superintegrable case}

A MS
  Hamiltonian is given by
\begin{equation*}
  {\cal H}^{\rm MS}_z =\frac 12  \jp^{(2)} {\rm e}^{ z \jm^{(2)}}=\frac12
\left( \frac {\sinh z
 q_1^2}{z q_1^2} \, {\rm e}^{z q_1^2}{\rm e}^{2z q_2^2}  p_1^2   +
 \frac {\sinh z q_2^2}{z q_2^2} \, {\rm e}^{ z q_2^2}  p_2^2  \right),
% \label{bi}
\end{equation*}
 since there exists  an additional (and functionally independent)
 constant of the motion~\cite{Deform}:
\begin{equation}
 {\cal I}_z=\frac {\sinh z
 q_1^2}{2 z q_1^2} \, {\rm e}^{z q_1^2}  p_1^2 .
 \label{bjjzz}
\end{equation}
 This choice corresponds to  the
 kinetic energy
\begin{equation*}
 {\cal T}^{\rm MS}_z(q_i,\dot q_i)=\frac 12 \left(\frac {z
 q_1^2}{\sinh z q_1^2} \, {\rm e}^{-z q_1^2}{\rm e}^{-2 z q_2^2} \,\dot
 q_1^2   +
 \frac {z q_2^2}{\sinh z q_2^2} \, {\rm e}^{-z q_2^2}\, \dot q_2^2
 \right)  ,
% \label{ea}
\end{equation*}
whose associated {metric} is
\begin{equation*}
 \d s_{\rm MS}^2=\frac {2z q_1^2}{\sinh z
 q_1^2} \, {\rm e}^{-z q_1^2}{\rm e}^{-2 z q_2^2} \,\d q_1^2   +
  \frac {2 z q_2^2}{\sinh z q_2^2} \, {\rm e}^{-z q_2^2} \, \d q_2^2 .
% \label{eecc}
\end{equation*}
Surprisingly enough, the computation of the Gaussian curvature $K$
for $\d s_{\rm MS}^2$ gives that $K=z$. Therefore, we are dealing
with a space of constant curvature which is just  the deformation
parameter $z$. In \cite{plb} it was shown that a certain change of
coordinates (that includes the signature parameter $\la_2$)
transforms the metric into
 \begin{equation*}
 \d s_{\rm MS}^2=
  \d r^2  +\la_2^2\,\frac{\sin^2(\la_1 r)}{\la_1^2} \, \d \te^2   ,
% \label{eh}
\end{equation*}
which exactly coincides with the metric of the Cayley--Klein
spaces written in geodesic polar coordinates
 $(r,\te)$   provided that now $z=\la_1^2\equiv \kappa_1$ and
 $\la_2^2\equiv
 \kappa_2$.
Obviously, after this change of variables the geodesic motion can
be reduced to a ``radial'' 1D system:
\begin{equation*}
 \widetilde{H}^{\Stc}_z=\frac 12 \, p_r^2
 +\frac{\la_1^2}{2\la_2^2\sin^2(\la_1 r)} \,
  \widetilde{C}_z  ,
 %\label{ffnn}
\end{equation*}
where $\widetilde{H}^{\Stc}_z=2{\cal H}^{\Stc}_z$ and
$\widetilde{C}_z=p_\te^2$ is, as in the previous case, the usual
generalized momentum for the $\te$ coordinate.

%%%%%%%%%%%%%%%%%%%%%%%%%%%%%%%%%%%%%%%%%%%%%%%%%%%%%%%%%%%%%%%%%%%%%%%%%%%%%%%%%

\subsubsection{A more general case}

At this point, one could wonder whether there exist other choices
for the Hamiltonian yielding constant curvature. In fact, let us
consider the generic Hamiltonian (\ref{bbfff}) depending on $f$.
If we compute the
 general expression for the  2D Gaussian curvature in terms of the function
$f(x)$ we f\/ind that
 \begin{gather*}
K(x)=z\left(f^\prime(x)\cosh x  +\left(
 f^{\prime\prime}(x)-f(x)-\frac{{f^\prime}^2(x)}{f(x)}
 \right) \sinh x
 \right),
 \end{gather*}
 where $x\equiv z\jm=z(q_1^2+q_2^2)$, $f^\prime=\frac{{\rm
 d}f(x)}{{\rm d}
 x}$ and $f^{\prime\prime}=\frac{{\rm d}^2f(x)}{{\rm d} x^2}$. Thus, in
general, we obtain spaces with variable curvature. In order to
characterize the constant curvature cases, we can def\/ine $g:=
f^\prime/f$ and write
\[
K/z = f^\prime \cosh x + \left(f^{\prime \prime} -f -
(f^\prime)^2/f \right) \sinh x  =  f\left( g \cosh x + (g^\prime
-1)\sinh x \right).
\]
If we now  require $K$ to be a constant we get the equation
\[
K^\prime=0 \equiv  2y\cosh x + y^\prime \sinh x = 0 , \qquad
\mbox{where}\quad y:=2g^\prime + g^2 -1 .
\]
The solution for this equation yields
\[
{{y=\frac{A}{\sinh^2 x}}},
\]
where $A$ is a constant, and solving for ${{g}}$, we  get for
${{F:=f^{\frac{1}{2}}}}$ the equation
\[
F^{\prime \prime}= \frac{1}{4} \left( 1+ {\frac{A}{\sinh
^2x}}\right)F ,
\]
whose  general solution is (${{A:=\lambda(\lambda-1)}}$):
\begin{equation*}
{{F= (\sinh x)^\lambda\left\{ C_1 \bigl(\sinh
(x/2)\bigr)^{(1-2\lambda)} + C_2\bigl( \cosh
(x/2)\bigr)^{(1-2\lambda)}\right\} }},
\end{equation*}
where $C_1$ and $C_2$ are two integration constants.

Therefore, many dif\/ferent solutions lead to 2D constant
curvature spaces. However,   we must impose as additional
condition that $\lim\limits_{x\to 0}{f}=1$. In this way   we
obtain that only the cases with $A=0$ are possible, that is,
either $\lambda=1$ or $\lambda=0$. Hence the two elementary
solutions   are just the Hamiltonians
\begin{equation*}
{\cal H}_z=\frac 12  \jp {\rm e}^{\pm  z \jm},
\end{equation*}
and the curvature of their associated spaces is $K=\pm z$.

%%%%%%%%%%%%%%%%%%%%%%%%%%%%%%%%%%%%%%%%%%%%%%%%%%%%%%%%%%%%%%%%%%%
\subsection{3D   curved manifolds}

The study of the 3D case follows exactly the same pattern. The
three-particle symplectic realization of  $sl_z(2,\mathbb R)$
(with all $b_i=0$) is obtained from (\ref{zsymp}):
\begin{gather*}
\jm^{(3)} = q_1^2+q_2^2+q_3^2\equiv \>q^2  ,\nonumber\\
 \jp^{(3)} =
  \frac {\sinh z q_1^2}{z q_1^2}\,  p_1^2\, {\rm e}^{z q_2^2}{\rm
e}^{z q_3^2} +
 \frac {\sinh z q_2^2}{z q_2^2} \, p_2^2  \, {\rm e}^{-z q_1^2} {\rm
e}^{z q_3^2}
 +
 \frac {\sinh z q_3^2}{z q_3^2} \, p_3^2 \,   {\rm e}^{-z q_1^2} {\rm
e}^{-z q_2^2} ,%\label{bees}
 \\
 \jj^{(3)} =
\frac {\sinh z q_1^2}{z q_1^2 }\,  q_1 p_1   {\rm e}^{z q_2^2}
{\rm e}^{z q_3^2} + \frac {\sinh z q_2^2}{z q_2^2 } \, q_2 p_2
{\rm e}^{-z q_1^2}{\rm e}^{z q_3^2}  + \frac {\sinh z q_3^2}{z
q_3^2 } \, q_3 p_3   {\rm e}^{-z q_1^2}{\rm e}^{-z q_2^2}
.\nonumber
\end{gather*}
By construction,  these generators  Poisson-commute with  the
three integrals $\{ { C}_z^{(2)},
 { C}_z^{(3)}\equiv C_{z,(3)}$, $C_{z,(2)}
\}$ given in (\ref{fk}):
\begin{gather}
  { C}_z^{(2)}    = \frac {\sinh z q_1^2 }{z q_1^2 } \,
\frac {\sinh z q_2^2}{z q_2^2} \left({q_1}{p_2} -
{q_2}{p_1}\right)^2 {\rm e}^{-z q_1^2}{\rm e}^{z
q_2^2} ,\nonumber\\
 C_{z,(2)} = \frac {\sinh z q_2^2 }{z q_2^2 } \, \frac
{\sinh z q_3^2}{z q_3^2} \left({q_2}{p_3} - {q_3}{p_2}\right)^2
{\rm e}^{-z q_2^2}{\rm e}^{z
q_3^2} ,\nonumber\\
 { C}_z^{(3)}   = \frac {\sinh z
q_1^2 }{z q_1^2 } \, \frac {\sinh z q_2^2}{z q_2^2}
\left({q_1}{p_2} - {q_2}{p_1}\right)^2 {\rm e}^{-z q_1^2}{\rm
e}^{z
q_2^2} {\rm e}^{2 z q_3^2} \label{3cas}\\
\phantom{{ C}_z^{(3)}   =}{}+ \frac {\sinh z q_1^2 }{z q_1^2 } \,
\frac {\sinh z q_3^2}{z q_3^2} \left({q_1}{p_3} -
{q_3}{p_1}\right)^2 {\rm e}^{-z q_1^2} {\rm e}^{  z
q_3^2}\nonumber  \\
\phantom{{ C}_z^{(3)}   =}{} + \frac {\sinh z q_2^2 }{z q_2^2 } \,
\frac {\sinh z q_3^2}{z q_3^2} \left({q_2}{p_3} -
{q_3}{p_2}\right)^2 {\rm e}^{-2z q_1^2}{\rm e}^{-z q_2^2} {\rm
e}^{z q_3^2}  . \nonumber
\end{gather}

%%%%%%%%%%%%%%%%%%%%%%%%%%%%%%%%%%%%%%%%%%%%%%%%%%%%%%%%%%%%%%%%
\subsubsection{QMS free motion: non-constant curvature}

If we now consider the {kinetic energy} ${\cal T}_z(q_i,\dot q_i)$
coming from  the Hamiltonian
\begin{equation}
{\cal H}_z(q_i,p_i)=\frac 12 \jp^{(3)}, \label{otroham}
\end{equation}
it corresponds to  the free Lagrangian~\cite{Checz}
\begin{equation*}
{\cal T}_z= \frac 12 \left(\frac
 {z q_1^2}{\sinh z q_1^2} \, {\rm e}^{-z q_2^2}{\rm e}^{-z q_3^2} \dot
q_1^2   +
 \frac {z q_2^2}{\sinh z q_2^2} \, {\rm e}^{z q_1^2}{\rm e}^{-z q_3^2}
\dot q_2^2    +
 \frac {z q_3^2}{\sinh z q_3^2} \, {\rm e}^{z q_1^2}{\rm e}^{z q_2^2}
\dot q_3^2
 \right)  ,
\end{equation*}
 that def\/ines a {geodesic f\/low on a 3D
 Riemannian space} with    metric
\begin{equation*}
\d s^2=\frac {2z q_1^2}{\sinh z
 q_1^2} \, {\rm e}^{-z q_2^2}{\rm e}^{-z q_3^2} \,\d q_1^2   +
  \frac {2 z q_2^2}{\sinh z q_2^2} \, {\rm e}^{z q_1^2}{\rm e}^{-z
q_3^2}\, \d q_2^2   +
  \frac {2 z q_3^2}{\sinh z q_3^2} \, {\rm e}^{z q_1^2}{\rm e}^{z
q_2^2}\, \d q_3^2 .
% \label{ccss}
\end{equation*}
The corresponding sectional curvatures $K_{ij}$ are
\begin{gather*}
K_{12}=\frac z4 \,{\rm e}^{-z \>q^2}\left( 1+ {\rm e}^{2 z q_3^2}-
2 {\rm e}^{2z
\>q^2}\right)  ,\\
K_{13}=\frac z4 \,{\rm e}^{-z \>q^2}\left( 2- {\rm e}^{2 z q_3^2}+
{\rm e}^{2 z q_2^2}{\rm e}^{2 z q_3^2}- 2 {\rm e}^{2z
\>q^2}\right)  ,\\
K_{23}=\frac z4 \,{\rm e}^{-z \>q^2}\left( 2-  {\rm e}^{2 z
q_2^2}{\rm e}^{2 z q_3^2}- 2 {\rm e}^{2z \>q^2}\right).
%\label{xa}
\end{gather*}
The following nice expression for the scalar curvature $K$ is
found:
\begin{equation*}
K_{12}+K_{13}+K_{23}=-\frac 52\,z\,\sinh(z\>q^2) = K/2.
%\label{xb}
\end{equation*}

Once again, the radial symmetry can be explicitly emphasized
through   new canonical coordinates $(\rho,\te,\tes)$ def\/ined
by:
\begin{gather}
\cosh^2(\la_1\rho)= {\rm e}^{2z \>q^2} ,\nonumber\\
\sinh^2(\la_1\rho)\cos^2(\la_2\te)={\rm e}^{2 z q_1^2}{\rm e}^{2 z
q_2^2}\left({\rm
e}^{2 z q_3^2}-1 \right)  ,\nonumber\\
 \sinh^2(\la_1\rho)\sin^2(\la_2\te)\cos^2\phi={\rm e}^{2 z
q_1^2}
\left({\rm e}^{2 z q_2^2}-1 \right)  ,\nonumber\\
  \sinh^2(\la_1\rho)\sin^2(\la_2\te)\sin^2\phi=  {\rm
e}^{2 z q_1^2}-1   , \label{xc}
\end{gather}
where $z=\la_1^2$ and $\la_2\ne 0$ is the additional signature
parameter, that will allow for  the presence of relativistic
spaces. Under this change of variables, the metric  is transformed
into
\begin{equation*}
 \d s^2=\frac {1}{\cosh(\la_1 \rr)}
 \left( \d \rr^2  +\la_2^2\,\frac{\sinh^2(\la_1 \rr)}{\la_1^2} \left(  \d
 \te^2 + \frac{\sinh^2(\la_2 \te)}{\la_2^2} \,\d\tes^2  \right) \right).
 %\label{xd}
\end{equation*}
This is just the  metric of the 3D Riemannian and relativistic
spacetimes written in geodesic polar coordinates and multiplied by
a global factor ${1}/{\cosh(\la_1 \rr)}$ that encodes the
information concerning the variable curvature of the space.

Sectional and scalar curvatures are now written in the form
\begin{equation*}
K_{12}=K_{13}= -\frac 12 \la_1^2 \,\frac{\sinh^2(\la_1
\rr)}{\cosh(\la_1
 \rr)},\qquad K_{23}=K_{12}/2 ,\qquad
K= -\frac 52 \la_1^2 \,\frac{\sinh^2(\la_1 \rr)}{\cosh(\la_1
 \rr)} .
\end{equation*}

Therefore, according to the values of $(\la_1,\la_2)$ we have
obtained a deformation of the 3D sphere $(i,1)$, hyperbolic
$(1,1)$, de Sitter   $(1,i)$ and anti-de Sitter $(i,i)$ spaces.
The ``classical" limit $z\to 0$   corresponds to a
zero-curvature limit leading to the proper
  Euclidean $(0,1)$ and Minkowskian $(0,i)$ spaces.

The QMS  Hamiltonian (\ref{otroham}), that determines the free
motion on the above spaces, and its three integrals of the motion
(\ref{3cas}) are written in terms of the new canonical coordinates
$(\rho,\te,\tes)$ and conjugated momenta $(p_\rho,p_\te,p_\tes)$
as{\samepage
\begin{gather}
{\tilde H}_z=\frac 12 {\cosh(\la_1 \rr)}
 \left( p_\rr^2  + \frac{\la_1^2}{\la_2^2\sinh^2(\la_1 \rr)} \left(
 p_\te^2 + \frac{\la_2^2}{\sin^2(\la_2 \te)}\,  p_\tes^2  \right) \right),
\nonumber\\
{\tilde C}_z^{(2)}=p_\tes^2, \qquad {\tilde C}_z^{(3)}=p_\te^2+
\frac{\la_2^2}{\sin^2(\la_2 \te)}\, p_\tes^2, \qquad {\tilde
C}_{z,(2)}=\left(\cos\tes\, p_\te-\la_2\frac{\sin\tes\,
p_\tes}{\tan(\la_2\te)}  \right)^2 , \label{mascas}
\end{gather}
provided that ${\tilde H}_z= 2{\cal H}_z$, ${\tilde  C}_z^{(2)}=4{
C}_z^{(2)}$, ${\tilde  C}_{z,(2)}= 4 \la_2^2 { C}_{z,(2)}$ and
${\tilde  C}_z^{(3)}= 4 \la_2^2 { C}_z^{(3)}$.}

Furthermore the  set of three functions $\{{\tilde
C}_z^{(2)},{\tilde C}_z^{(3)},{\tilde H}_z\}$,   which
characterizes the  complete integrability of the Hamiltonian,
allows us to
 write  three equations, each of them depending on a canonical pair:
\begin{gather}
{\tilde  C}_z^{(2)}(\phi,p_\phi)=p_\phi^2  ,\qquad {\tilde
C}_z^{(3)}(\theta,p_\theta)=p_\theta^2+\frac{\la_2^2}{
\sin^2(\la_2\theta)}\,{\tilde  C}_z^{(2)}
,\nonumber\\
{\tilde H}_z(\rr,p_\rr)=\frac 12 {\cosh(\la_1 \rr)}
 \left(   p_\rr^2+
\frac{\la_1^2}{\la_2^2 \sinh^2(\la_1 \rr) } \,{\tilde  C}_z^{(3)}
\right). \label{fee}
\end{gather}
Therefore the Hamiltonian is separable and  reduced to a 1D radial
system.

%%%%%%%%%%%%%%%%%%%%%%%%%%%%%%%%%%%%%%%%%%%%%%%%%%%%%%%%%%%%%%%%
\subsubsection{MS free motion: constant curvature}

The following choice for the Hamiltonian
\begin{equation*}
{\cal H}_z^{\rm MS}=\frac 12 \jp^{(3)} \,{\rm e}^{ z \jm^{(3)}},
\end{equation*}
yields a MS system  since it has four (functionally independent)
constants of motion, the three universal  integrals (\ref{3cas})
together with   ${\cal I}_z$ (\ref{bjjzz}). In fact, this the 3D
version of the Hamiltonian described in Section 4.1.2.

The associated {kinetic energy} is
\begin{equation*}
{\cal T}^{\rm MS}_z=\frac 12 \left(\frac
 {z q_1^2}{\sinh z q_1^2} \, {\rm e}^{z q_1^2}  \dot
q_1^2   +
 \frac {z q_2^2}{\sinh z q_2^2} \, {\rm e}^{2z q_1^2}{\rm e}^{z q_2^2}
\dot q_2^2    +
 \frac {z q_3^2}{\sinh z q_3^2} \, {\rm e}^{2z q_1^2}{\rm e}^{2z q_2^2}{\rm e}^{z
q_3^2} \dot q_3^2
 \right)  ,
% \label{ya}
\end{equation*}
and the underlying metric reads
\begin{equation}
\d s_{\rm MS}^2=\frac {2z q_1^2}{\sinh z
 q_1^2} \, {\rm e}^{z q_1^2}  \,\d q_1^2   +
  \frac {2 z q_2^2}{\sinh z q_2^2} \, {\rm e}^{2z q_1^2}{\rm e}^{z
q_2^2}\, \d q_2^2   +
  \frac {2 z q_3^2}{\sinh z q_3^2} \, {\rm e}^{2 z q_1^2}{\rm e}^{2 z
q_2^2}{\rm e}^{z q_3^2}\, \d q_3^2 .
 \label{yb}
\end{equation}
This space is again a {Riemannian} one with
 constant sectional and scalar curvatures given by
\begin{equation*}
K_{ij}=z ,\qquad K=6z .
%\label{yc}
\end{equation*}
Through an appropriate change of coordinates~\cite{Checz} we
f\/ind that  (\ref{yb}) is transformed into the 3D Cayley--Klein
metric written in terms of   geodesic polar coordinates
$(r,\te,\tes)$:
\begin{equation*}
 \d s^2_{\rm MS}= \d r^2  +\la_2^2\,\frac{\sin^2(\la_1 r)}{\la_1^2} \left(  \d
 \te^2 + \frac{\sin^2(\la_2 \te)}{\la_2^2} \,\d\tes^2  \right)   .
 %\label{ye}
\end{equation*}
Therefore, according to the values of  $(\la_1,\la_2)$, this
metric provides  the  3D
 sphere $(1,1)$,
Euclidean $(0,1)$, hyperbolic   $(i,1)$, anti-de Sitter $(1,i)$,
Minkowskian $(0,i)$, and de Sitter $(i,i)$ spaces.

Now the MS Hamiltonian, ${\tilde H}_z^{\rm MS}=2{\cal H}_z^{\rm
MS}$, is written as
\begin{equation*}
{\tilde H}_z^{\rm MS}=\frac 12
 \left( p_r^2  + \frac{\la_1^2}{\la_2^2\sin^2(\la_1 r)} \left(
 p_\te^2 + \frac{\la_2^2}{\sin^2(\la_2 \te)}\,  p_\tes^2  \right) \right),
 %\label{yf}
\end{equation*}
and the four functionally independent integrals are given by
(\ref{mascas}) and
\begin{equation*}
{\tilde I}_z =\left(\la_2\sin(\la_2\te)\sin\tes\,
p_r+\frac{\la_1\cos(\la_2\te)\sin\tes}{\tan(\la_1 r)}\,p_\te +
\frac{\la_1\la_2\cos \tes }{\tan(\la_1 r)\sin(\la_2\te)}\,p_\tes
\right)^2 ,
\end{equation*}
where  ${\tilde I}_z =4\la_2^2{\cal I}_z$. The {two sets}
$\{{\tilde H}_z^{\rm MS}, {\tilde C}_z^{(2)}, {\tilde
C}_z^{(3)}\}$ and $\{{\tilde H}_z^{\rm MS}, {\tilde C}_{z,(2)},
{\tilde I}_z\}$ consist of three functions in involution.
Similarly to (\ref{fee}),    this Hamiltonian  is also separable
and can be reduced to a~1D system:
 \begin{equation*}
{\tilde H}_z^{\rm MS}(r,p_r)=\frac 12
 \left( p_r^2  + \frac{\la_1^2}{\la_2^2\sin^2(\la_1 r)} \,{\tilde C}_z^{(3)} \right).
\end{equation*}

%%%%%%%%%%%%%%%%%%%%%%%%%%%%%%%%%%%%%%%%%%%%%%%%%%%%%%%%%%%%%%%%%%%
\section[$N$D spaces with variable curvature]{$\boldsymbol{N}$D spaces with variable curvature}

The generalization to arbitrary dimension is obtained through the
same procedure, and the starting point is the QMS Hamiltonian for
the $N$D  geodesic motion that, in the simplest case, reads:
\begin{equation*}
 {\cal H}_z = \frac 12
 \jp^{(N)}=\frac 12
 \sum_{i=1}^N
  \frac {\sinh z q_i^2}{z q_i^2}\,  p_i^2    \exp{\left( - z
 \sum_{k=1}^{i-1}  q_k^2+  z \sum_{l=i+1}^N   q_l^2 \right)} .
% \label{ddr}
\end{equation*}
The
 geometric characterization of the underlying $N$D curved
 spaces follows the same path as in the 2D and 3D cases described in the
previous sections.

If we write the above Hamiltonian as
\begin{equation*}
{ {\cal{H}}_z = \frac{1}{2} \sum_{i=1}^N s_z(q_i^2)\,p_i^2\exp
\left( z\sum_{   k=1; k  \neq i}^N  {\rm{sgn}}(k-i)q_k^2 \right)
},
\end{equation*}
where $s_z(q_i^2)=   {\sinh z q_i^2}/(z q_i^2)$ and ${\rm
sgn}(k-i)$ is the sign of the dif\/ference $k-i$, we get again a~free Lagrangian:
\begin{equation*}
{ {\cal{T}}_z = \frac{1}{2} \sum_{i=1}^N \frac{(\dot q_i)^2
\exp\left(-z\sum\limits_{   k=1; k  \neq i}^N
{\rm{sgn}}(k-i)q_k^2\right)}{s_z(q_i^2)}} ,
\end{equation*}
with the    corresponding (diagonal) metric given by
\begin{equation}
 \d s^2 = \sum_{i=1}^N g_{ii}(q) \, \d q_i^2 ,\qquad
 g_{ii}(q)=\frac{\exp\left(-z\sum\limits_{   k=1; k  \neq i}^N
{\rm{sgn}}(k-i)q_k^2\right)}{s_z(q_i^2)}. \label{mmetric}
\end{equation}

It turns out that the most suitable way to understand the nature
of the problem as well as to enforce separability is to consider
two sets of new coordinates:

\begin{itemize}\itemsep=0pt

 \item[$\bullet$] $N+1$ ``collective" variables~\cite{BRcluster} $({\xi _0},{\xi
_1},\dots,{\xi _N})$. They play a similar role to the ambient
coordinates arising when $N$D Riemannian spaces of constant
curvature are embedded within ${\mathbb R}^{N+1}$.

 \item[$\bullet$]  $N$ ``intrinsic" variables  $(\rho,\te_2,\dots,\te_N)$ which describe
the $N$D space itself. They   are the analogous to the geodesic
polar coordinates on $N$D Riemannian spaces of constant
curvature~\cite{VulpiLett,CRMVulpi}.

 \end{itemize}

The above coordinates are def\/ined in terms of the initial $q_i$
by:
\begin{gather}
{{\xi_0^2}} ={{ \cosh^2 (\la_1\rho):= \prod_{i=1}^N \exp( 2z q_i^2)}} ,\nonumber\\
{{\xi_k^2}}= {{ \sinh^2 (\la_1\rho)\prod_{j=2}^{k}\sin^2\theta_j
\cos^2 \theta_{k+1} := \prod_{i=1}^{N-k} \exp (2z
q_i^2)\left(\exp(2zq_{N-k+1}^2)
-1\right)}} ,\label{changen}\\
{{\xi_N^2}}={{\sinh^2(\la_1\rho)\prod_{j=2}^N\sin^2\theta_j
}}:=\exp(2zq_{1}^2) -1, \nonumber
\end{gather}
 where $z=\la_1^2$, $k=1,\dots,N-1$, and hereafter a product $\prod_j^k$ such that $j>k$
is assumed to be equal to 1. Notice also that for the sake of
simplicity we have not introduced the additional signature
parameter $\la_2$ (which would have been associated with $\te_2$).
This def\/inition is the $N$D generalization of the   change of
coordinates (\ref{xc}) given in the 3D case with $\te=\theta_2$,
$\tes=\theta_3$ and $\la_2=1$.

Clearly, the $N+1$ collective variables are not independent and
they fulf\/il a pseudosphere relation (of hyperbolic type):
\begin{equation*}
\xi _0^2-\sum_{k=1}^N \xi _k^2 = 1.
\end{equation*}

The geodesic f\/low in the canonical coordinates  ${{(\rho,
\theta)}}$ and momenta ${{(\rho,  p_\theta)}}$ is then given by
the Hamiltonian ${\tilde {H}}_z=2 {\cal{H}}_z$:
\begin{gather*}
{{ {\tilde{H}}_z  = \frac 12 \cosh(\la_1\rho) \left( p_{\rho}^2
+\frac{\la_1^2}{\sinh^2 (\la_1\rho)} \sum_{i=2}^N
\left(\prod_{j=2}^{i-1}\frac{1}{\sin^2
\theta_j}\right)p_{\theta_i}^2\right) }},
\end{gather*}
and the (left) integrals of the motion $\tilde C^{(m)}_z= 4
C^{(m)}_z$ are found to be
\begin{gather*}
\tilde C^{(m)}_z=\sum_{i=N-m+2}^N    \left(\prod_{j=N-m+2}^{i-1}
\frac{1}{\sin^2 \theta_j}\right)p_{\theta_i}^2,\qquad m=2,\dots,N
.
\end{gather*}
By taking into account the $N$ functions $\{{\tilde{H}}_z,\tilde
C^{(m)}_z\}$, we obtain the following set of $N$ equations, each
of them depending on a single canonical pair, which shows the
reduction of the system to  a 1D problem:
\begin{gather*}
\tilde C^{(2)}_z(\te_N,p_{\te_N})=p_{\te_N}^2,\nonumber\\
\tilde
C^{(m)}_z(\te_{N-m+2},p_{\te_{N-m+2}})=p_{\te_{N-m+2}}^2+\frac{
C^{(m-1)}_z}{\sin^2\te_{N-m+2}},\qquad m=3,\dots,N,\nonumber\\
{\tilde{H}}_z(\rho,p_\rho) = \frac 12 \cosh(\la_1\rho) \left(
p_{\rho}^2 +\frac{\la_1^2}{\sinh^2 (\la_1\rho)}  \,\tilde
C^{(N)}_z  \right).
\end{gather*}
We stress that these models can be extended to incorporate
appropriate  interactions with an external central f\/ield,
preserving superintegrability. This will be achieved by modifying
the Hamiltonian by adding an arbitrary function of $J_-$, as we
shall see in the next Section.

Finally we remark that the corresponding generalization to $N$D
spaces with constant curvature can be obtained by considering the
MS Hamiltonian
\begin{equation}
 {\cal H}^{\rm MS}_z = \frac 12 \,
 \jp^{(N)}\,{\rm e}^{ z \jm^{(N)}}  =  \frac 12\,{\rm
e}^{ z \>q^2}
 \sum_{i=1}^N
  \frac {\sinh z q_i^2}{z q_i^2}\,  p_i^2    \exp{\left( - z
 \sum_{k=1}^{i-1}  q_k^2+  z \sum_{l=i+1}^N   q_l^2 \right)} .
\label{msham}
\end{equation}
Under a suitable change of coordinates, similar to (\ref{changen})
but involving  a dif\/ferent radial coordinate~$r$ instead of
$\rho$, this Hamiltonian leads to the MS geodesic motion on ${\>
S}^N$, ${\> H}^N$ and ${\> E}^N$ in the proper geodesic polar
coordinates which can be found in~\cite{VulpiLett,CRMVulpi}.

\section{QMS potentials}

As we have just noticed, we can also consider more general $N$D
QMS Hamiltonians based on $sl_z(2,{\mathbb R})$ (\ref{zsymp}) by
considering arbitrary $b_i$'s (contained in $J_+$) and adding some
functions depending on $J_-$; hereafter we drop the index ``$(N)$"
in the generators. The family of Hamiltonians that we consider has
the form (see \cite{jpa2D} for the 2D construction):
\begin{equation*}
{\cal H}_z=\frac 12 \jp \, f (z\jm
 )+\pot (z\jm )  ,
%\label{ahaa}
\end{equation*}
where the arbitrary smooth functions $f$ and $\cal U$ are such
that
\begin{equation*}
\lim_{z\to 0}\pot(zJ_-)={\cal V}(J_-), \qquad \lim_{z\to
0}f(zJ_-)=1.
\end{equation*}
This, in turn, means that
\begin{equation*}
\lim_{z\to 0} {\cal H}_z= \frac 12 \,\>p^2 + {\cal V}(\>q^2)
+\sum_{i=1}^N\frac{b_i}{2q_i^2},
\end{equation*}
recovering the superposition of a central potential ${\cal
V}(J_-)\equiv {\cal V}(\>q^2)$ with $N$ centrifugal terms on~${\>
E}^N$~\cite{Evansa}. Such a ``f\/lat" system has a (non-deformed)
$sl(2,{\mathbb R})$ coalgebra symmetry as given in Theorem 1.

We recall that the function $f (z\jm
 )$ gives us the type of
{curved background}, which is characterized by the metric  $\d
s^2/f (z\>q^2 )$ where $\d s^2$ is the variable curvature metric
associated to ${\cal H}_z=\frac 12 \jp$ and given in
(\ref{mmetric}). The two special cases with $f (z\jm
 )={\rm e}^{\pm z J_-}$ give rise to Riemannian spaces of constant sectional
curvatures, all equal to $\pm z$ (as (\ref{msham})).

In particular,  QMS deformations of the
 $N$D  SW
system would be given by any  $\pot$ such that
\begin{equation*}
\lim_{z\to 0} \pot(zJ_-) = \omega\jm ,
\end{equation*}
and for the $N$D generalized KC potential we can consider  $\pot$
functions such that
\begin{equation*}
\lim_{z\to 0} \pot(zJ_-)= -k/\sqrt{\jm}.
\end{equation*}
In both cases   centrifugal type potentials come from the $b_i$'s
terms contained in $J_+f (z\jm  )$.

With account of the geometrical arguments, the following QMS SW
system on  spaces with non-constant curvature (\ref{mmetric}) has
been proposed in \cite{jpa2D}:
\begin{equation*}
{\cal H}^{\rm SW}_z=\frac 12 \jp +
 {\omega} \,\frac{\sinh{z \jm}}{z},
\end{equation*}
while a   candidate for a generalized {KC system on such spaces}
is given by the formula \cite{jpa2D}:
\begin{equation*}
{\cal H}^{\rm KC}_z=\frac 12 \jp -
 k \,\sqrt{  \frac{2z}{{\rm e}^{2 z \jm}-1}} \,{\rm e}^{2 z \jm} .
\end{equation*}

In the constant curvature case, the approach here presented allows
us to recover the known results for the SW potential on Riemannian
spaces with constant curvature, as well as their generalization to
relativistic spaces (whenever the signature parameter $\la_2$ is
considered).

In particular,  the MS {SW system on $N$D spaces of constant
curvature} is given by the Hamiltonian
\begin{equation*}
 {\cal H}^{\rm MS,SW}_z=\frac 12 \jp {\rm e}^{ z \jm}+
 {\omega} \,\frac{\sinh{z \jm}}{z} \,{\rm e}^{ z \jm}\equiv {\cal H}^{\rm
SW}_z   {\rm e}^{ z \jm} ,
%\label{bh}
\end{equation*}
and the additional constant of the motion that provides the MS
property reads
\begin{equation*}
{\cal I}_z=\frac {\sinh z q_1^2}{2 z q_1^2} \, {\rm e}^{z q_1^2}
p_1^2 +\frac{z \otra_1}{2\sinh z q_1^2} \, {\rm e}^{z
q_1^2}+\frac{\omega}{2z}\,{\rm e}^{2 z q_1^2} .
%\label{bjj}
\end{equation*}

The corresponding results for the generalized KC system are
currently under investigation.

%%%%%%%%%%%%%%%%%%%%%%%%%%%%%%%%%%%%%%%%%%%%%%%%%%%%%%%%%%%%%%%%%%%
\section{Concluding remarks}

The main message that we would like to convey to the scientif\/ic
community through the present paper is that  ``Superintegrable
Systems are not rare!". Indeed, in our approach they turn out to
be a natural manifestation of coalgebra symmetry: as such, they
can be equally well constructed on a f\/lat or on a curved
background, the latter being possibly equipped with a variable
curvature. Moreover, and, we would say,  quite remarkably the
construction holds for an arbitrary number of dimensions.

In that perspective, the most interesting  problems that are still
open are in our opinion the following ones:

\begin{enumerate}\itemsep=0pt
\item The explicit integration of  the equations of motion for (at
least some) of the prototype examples we have introduced in the
previous sections; \item The construction of the
quantum-mechanical counterpart of  our approach.

\end{enumerate}

\noindent As for the former point, partial  results have already
been obtained, and a detailed description of the most relevant
examples will be published soon. The latter point, in particular
as far as  the non-standard deformation of $sl(2,\mathbb R)$ is
concerned, is however  more subtle and deserves careful
investigation (which is actually in progress). In fact, f\/irst of
all  one has to f\/ind a proper $\infty$-dimensional
representation of such a non-standard deformation in terms of
linear operators acting on a suitably def\/ined Hilbert space,
ensuring self-adjointness of the Hamiltonians; second, and
certainly equally important, at least in some physically
interesting special cases one would like to exhibit the explicit
solution of the corresponding spectral problem.

%%%%%%%%%%%%%%%%%%%%%%%%%%%%%%%%%%%%%%%%%%%%%%%%%%%%%%%%%%%%%%%%%%%%%%%%%%%%%%%%%

\subsection*{Acknowledgements}

This work was partially supported  by the Ministerio de
Educaci\'on y Ciencia   (Spain, Project FIS2004-07913),  by the
Junta de Castilla y Le\'on   (Spain, Project VA013C05), and by the
INFN--CICyT (Italy--Spain).

\pdfbookmark[1]{References}{ref}
\LastPageEnding

\end{document}